\documentclass[twocolumn]{aastex631}
\usepackage{multirow}
\usepackage{amsmath}
\usepackage{graphicx}
\usepackage{threeparttablex}  

\newcommand{\ecyc}{E$_{\texttt{cyc}}$} 
\newcommand{\Lx}{$L_{\texttt{X}}$}

\newcommand{\astrosat}{\textit{AstroSat}}
\newcommand{\hxmt}{{\textit{Insight}}-HXMT}
\newcommand{\nustar}{\textit{NuSTAR}}

\begin{document}

\title{Discovery of a new phase-transient cyclotron line in A0535+26: Constraints on the accretion geometry}

\author{Ashwin Devaraj}
\affiliation{Raman Research Institute,  
Sadashivanagar\\ Bangalore 560080, India\\}

\affiliation{Joint Astronomy Programme, 
Indian Institute of Science\\
Bangalore 560012, India\\}

\author{Biswajit Paul}
\affiliation{Raman Research Institute, 
Sadashivanagar\\ Bangalore 560080, India\\}

\author{Varun Bhalerao}
\affiliation{Indian Institute of Technology Bombay, 
Powai\\ Mumbai 400076, India\\}

\author{Dipanjan Mukherjee}
\affiliation{ Inter-University Centre for Astronomy and Astrophysics, 
\\ Pune  411007, India\\}

\correspondingauthor{Ashwin Devaraj}
\email{devarajashwin@gmail.com}



\begin{abstract}

In November 2020, A0535+26 underwent one of its brightest outbursts, reaching nearly 12 Crab in X-ray flux. Observed by \hxmt, \nustar, \textit{NICER}, and \astrosat, this event provided valuable insights into Be/X-ray binaries. The pulse profiles evolved significantly with luminosity, transitioning from pencil-beam to fan-beam geometries. A0535+26, known for its fundamental cyclotron line at $\sim$44 keV, became only the second source to exhibit a negative correlation between cyclotron line energy and flux at high luminosities, with a plateau phase preceding the transition from positive to negative correlation. We report the discovery of a phase-transient low-energy cyclotron line, detected in a narrow phase range ($\sim$16\%) across all seven \nustar\ observations during the rising, peak, and declining phases of the outburst. The new line exhibited dramatic variations with pulse phase and luminosity. We explain this behavior using an accretion geometry where the accretion column sweeps across the line of sight.

\end{abstract}



\section{Introduction} \label{sec:intro}

Neutron stars (NS) are highly compact objects with some of the strongest magnetic fields in the universe, ranging from 10$^8$ to 10$^{15}$ G. In most Be/X-ray binaries, a NS orbits a massive B-type star in an eccentric orbit. These stars emit X-rays as the NS accretes material from a disk outflowing from its fast-rotating companion \citep{Reig_2011}. The accreted matter halts at the magnetospheric radius and is channeled onto the NS’s magnetic poles, producing X-ray pulsations when the magnetic and spin axes are misaligned.

The most direct way to probe the magnetic fields of NS is through the detection of Cyclotron Resonant Scattering Features (CRSF) or cyclotron lines. These are absorption features in the hard X-ray spectrum of accreting pulsars, typically modeled using Lorentzian or Gaussian absorption profiles. The centroid energy of these features relates to the magnetic field through $E_{\mathrm{cyc}} \sim 11.6 \ n \ \mathrm{B}_{12}$ keV, where B$_{12}$ is the magnetic field strength in units of $10^{12}$ G, and $n$ is the harmonic (see \citealt{Staubert_2019} or \citealt{Maitra_2017} for a review on cyclotron lines in neutron stars). CRSF parameters are known to vary over time \citep{staubert_2014}, with X-ray luminosity \citep{Rothschild_2017, Tsygankov_2006}, and with the neutron star’s rotational phase \citep{Suchy_2008, Varun_2019a}. Luminosity-dependent variations in cyclotron lines are used to study accretion regime transitions in NS. 

A0535+26 was discovered in 1975 with Ariel V with a spin period of 104 seconds \citep{Rosenberg_1975}. It was later found to have an orbital period of about 111 days \citep{Priedhorsky_1983} and is located approximately 2 kpc away \citep{Steele_1998}. Although a transient source, A0535+26 has exhibited several outbursts since its discovery and has been extensively studied. It is one of the few sources observed at the lower range of luminosities, around $10^{34-35}$ erg s$^{-1}$ and the highest luminosities reaching up to $10^{38}$ erg s$^{-1}$. At low luminosities, the energy spectrum is described by two power-laws with cutoffs rather than the typical single power-law \citep{Tsygankov_2017}.  At high luminosities, the spectral shape can be described with a power law modified by an exponential cutoff and the pulse profile undergoes dramatic shape changes \citep{Mandal_2022, Xiao_2024}. A cyclotron line at 45 keV and its harmonic at 90 keV were detected, indicating a magnetic field of about $4 \times 10^{12}$ G \citep{Caballero_2007}. A 0535+26 is the only source that exhibits all three types of luminosity-cyclotron correlations: no correlation at the lowest luminosities \citep{Caballero_2007}, a positive correlation over a range of intermediate luminosities \citep{Klochkov_2011, Muller_2013, Sartore_2015}, and a negative correlation at high luminosities \citep{Kong_2021}. 

\begin{figure}
    \centering
 \vspace{0.5cm}
	\includegraphics[scale=0.35,trim={0 1.0cm 0 2.0cm},  angle=-90]{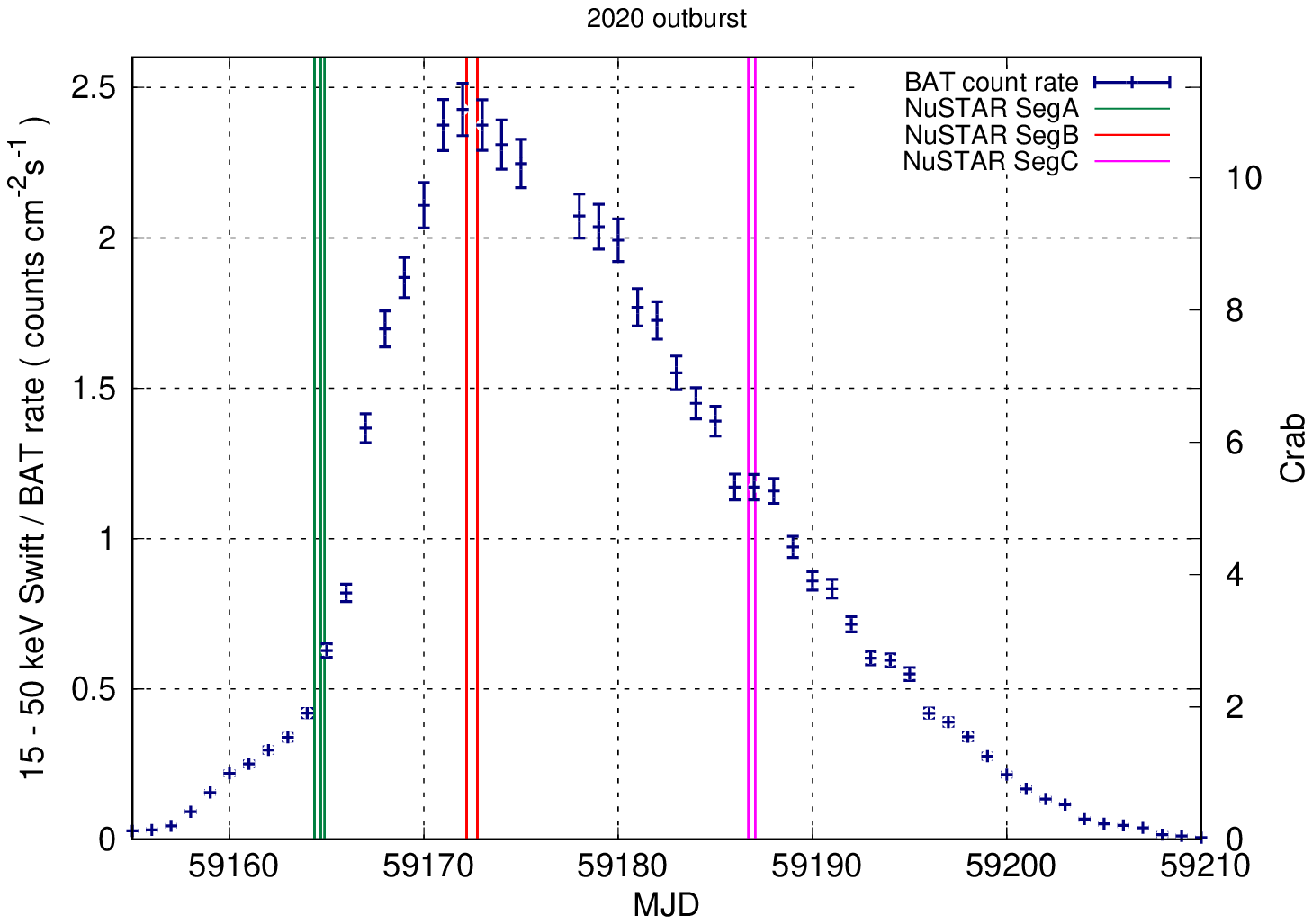}
    \caption{\textit{Swift/BAT} light curve of the 2020 outburst. Blue points indicate the \textit{Swift/BAT} count rate while the three green vertical lines indicate the times of the first set (\textbf{SegA}) of \textit{NuSTAR} observations, red the second set (\textbf{SegB}), and magenta the third set (\textbf{SegC}) respectively. The data were obtained from \url{https://swift.gsfc.nasa.gov/results/transients/1A0535p262/}}
    \label{fig:outburst-lightcurve}
\end{figure}

Around November 2020, A0535+26 experienced its brightest outburst since its discovery, lasting about 50 days and reaching a peak flux of nearly 12 Crab (see Fig. \ref{fig:outburst-lightcurve}). This outburst, observed with many observatories, provided key insights into the nature of this X-ray pulsar. Optical studies of the $H \alpha$ line profile evolution suggested that mass accretion from a warped disk may have triggered the event \citep{Chhotaray_2023}. \citet{Kong_2022} conducted a detailed study of the pulse-phase dependence of the cyclotron line and its first harmonic, finding evidence of photon spawning, where the optical depth of the harmonic exceeds that of the fundamental, making the fundamental feature shallower \citep{Schonherr2007}. The pulse profile evolution, explored by \citet{Wang_2022} and \citet{Xiao_2024}, showed significant luminosity-dependent changes, pointing to an evolving X-ray emission geometry. Pulse-profile decomposition by \citet{Hu_2023} indicated emission from two poles misaligned by about 12$^{\circ}$ from being antipodal, with transitions between “pencil” and “fan” beam geometries. Using \hxmt \ and \nustar \ data, \citet{Kong_2021} and \citet{Mandal_2022} found a negative correlation between cyclotron line energy and luminosity, while \citet{Shui_2024} discovered complex variations in the cyclotron line, showing an initial increase with luminosity that plateaus before decreasing.

The \nustar \ telescope was used to observe this source seven times during the outburst: three times during the rising phase, twice at the peak, and twice during the decline (see Fig.\ref{fig:outburst-lightcurve}). The timing and phase-average spectral study of these observations were performed by \citet{Mandal_2022}. In this work, using these \nustar \ observations, we carried out pulse phase resolved spectroscopy and report the discovery of a new phase-transient cyclotron line. We discuss the implications of the detection of such a line and suggest an accretion geometry for this system that accounts for the observational findings.

\section{Observations and data reduction} \label{sec:data-red}

The Nuclear Spectroscopic Telescope Array (\nustar) is a hard X-ray telescope operating in the 3–79 keV range, designed to study high-energy cosmic phenomena \citep{Harrison_2013}. It uses grazing incidence optics to focus X-rays onto two identical focal plane modules, FPMA and FPMB, each equipped with four solid-state CdZnTe detectors. With an energy resolution of about 0.4 keV at 10 keV and 0.9 keV at 60 keV, \textit{NuSTAR} is ideal for studying cyclotron line sources.

The details of the observations used in this work are listed in Table \ref{tab:nustar_obs}. \textbf{SegA} refers to the rising phase of the outburst, \textbf{SegB} to the peak, and \textbf{SegC} to the declining phase. The data was processed using the standard analysis suite \textsc{heasoft} v6.33c and \textsc{nustardas} v2.1.2 (CALDB version: 20230918). Circular regions of about 140 arcsec were chosen for the source and background due to the source’s brightness. The event files were Solar System barycenter corrected, and the \texttt{nuproducts} script was used to extract the spectrum and light curve for both FPMA and FPMB in the standard way.

\begin{table*}[ht]
\centering
\begin{tabular}{|c|c|c|c|c|c|}
\hline
\textbf{Segment} & \textbf{Obs Number} & \textbf{OBS\_ID} & \textbf{Obs start time} & \textbf{Obs duration (ks)} & \textbf{Avg \textit{NuSTAR} Count Rate} \\ \hline
\textbf{SegA} & obs1 & 90601334002 & 2020-11-11T08:29:36 & 20.7 & 669 \\ 
            & obs2 & 90601334003 & 2020-11-11T16:32:43 & 9.1  & 751 \\ 
            & obs3 & 90601334004 & 2020-11-11T21:22:35 & 20.7 & 757 \\ \hline
\textbf{SegB} & obs4 & 90601335002 & 2020-11-19T04:54:31 & 26.3 & 3263 \\ 
            & obs5 & 90601335003 & 2020-11-19T17:47:33 & 9.2  & 3241 \\ \hline
\textbf{SegC} & obs6 & 90601336002 & 2020-12-03T16:47:53 & 20.7 & 1826 \\ 
            & obs7 & 90601336003 & 2020-12-04T00:51:10 & 14.9 & 1846 \\ \hline
\end{tabular}
\caption{Observation Details for \textit{NuSTAR} Observations}
\label{tab:nustar_obs}
\end{table*}

\section{Analysis and results} \label{sec:floats}
\subsection{Timing}
\begin{figure}
    \centering
 \vspace{0.5cm}
	\includegraphics[angle=90,scale=0.5,trim={0 0.0cm 0cm 0 2.0cm},  angle=-90]{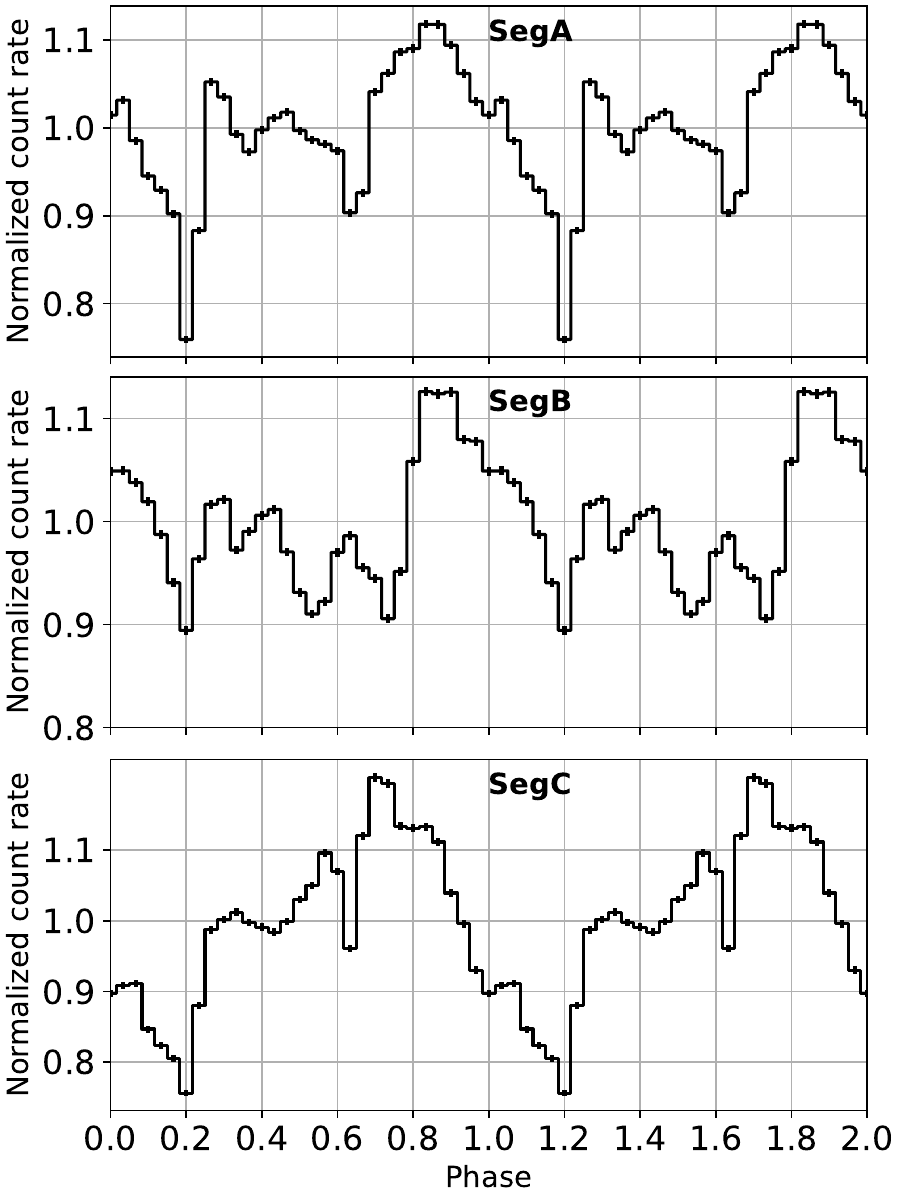}
    \caption{Pulse profiles of the lightcurves in \textbf{SegA, SegB,} and \textbf{SegC} in the 3-79 keV band from the top to bottom. Each pulse profile has been constructed by folding the combined lightcurves within each segment.}
    \label{fig:pulse-prof-allseg}
\end{figure}
The timing results for all \nustar \ observations were presented by \citet{Mandal_2022}, where the source was found to be spinning up, and its pulse profile showed clear energy dependence. The pulse profiles show complex shapes that vary across different energy bands and evolve with luminosity. \citet{Wang_2022} studied the evolution of the pulse profiles across the 2020 outburst using \textit{Insight-HXMT} in multiple energy bands while \citet{Xiao_2024} looked at the pulse profiles before and after the outburst using data from \textit{NICER}. One of the common features in all the pulse profiles of this object, including the multiple observations from previous outbursts, is the presence of a characteristic deep minima that persists irrespective of the evolution of the pulse profiles (see Fig.2 in \citealt{Wang_2022} and Fig.2 in \citealt{Maitra2013}).  

 In our work, since the observations in each segment were performed close together, we combined the light curves for each segment and estimated the periods as P${_\mathrm{segA}} =$ 103.564(6), P${_\mathrm{segB}} =$ 103.455(4), and P$_{\mathrm{segC}} =$ 103.266(2). We then generated the pulse profiles using the combined lightcurves for each of the three segments (see Fig.\ref{fig:pulse-prof-allseg}). We aligned the characteristic dip of the there segments to the phase 0.2 by comparing the pulse profiles with the  evolution of the \hxmt \  pulse profiles from \citet{Wang_2022}. Apart from the main dip at the 0.2 phase, there are other minima at phases 0.65 and around 1.0 which show changes with luminosity. 

 \begin{figure*}
    \centering
	\includegraphics[scale=0.3, trim={0 3.0cm 0 1.8cm},  angle=-90]{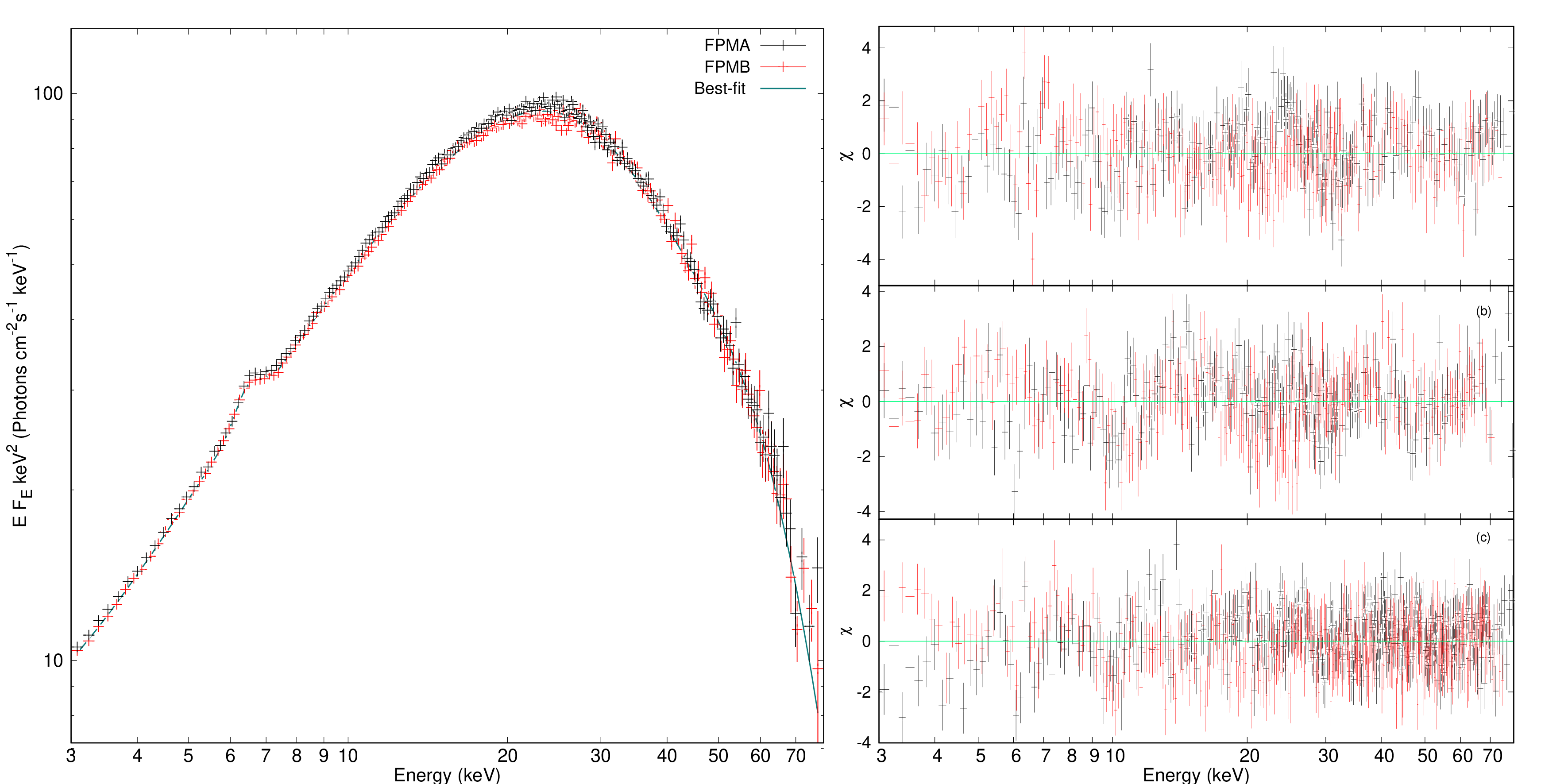}
    \caption{ Phase-average spectra: The unfolded spectrum with the best-fit of \texttt{FDCut} model is shown on the left panel with the spectrum of obs3. The right panel shows the residuals. Panel (a) best fit for obs3 in \textbf{SegA} (b) best fit  for obs5 in \textbf{SegB}  (c) best fit for obs7 in \textbf{SegC} . The red and black points correspond to data from FPMA and FPMB respectively.}
    \label{fig:unfolded-avg-spectra-residues} 
\end{figure*}
\subsection{Phase average spectroscopy}

For all observations, the spectra from both modules were optimally binned following the method prescribed by \citet{Kaastra_2016}. The spectra were fit simultaneously, allowing for a relative normalization constant, $C_B$, while fixing the FPMA normalization to unity. Typically, we found $C_B \sim 0.98$. The phase-average model from \citet{Mandal_2022} was “\texttt{phabs} $\times$ (\texttt{cutoffpl} + \texttt{gauss} + \texttt{bbody}) $\times$ \texttt{gabs}$_{\mathrm{cyc}}$,” representing low-energy interstellar absorption, a cutoff power-law for the continuum, a Gaussian emission line for the Fe K$\alpha$, a blackbody, and a Gaussian absorption profile for the 44 keV cyclotron line. They reported a high blackbody temperature of $kT_{bb} \sim 6.3$ keV, implying a peak temperature around 16 keV, which would significantly contribute to the hard X-ray continuum. Since such a high-temperature blackbody component is rare, we explored other models for the hard X-ray continuum and the possible presence of a low-temperature blackbody component. Two alternate models that have been commonly used are: the \texttt{FDCut}—with a Fermi-Dirac-like cutoff offering a smooth transition at the cutoff energy \citep{Tanaka_1986}—and the \texttt{NPEX} model, which combines two cutoff power-laws with a common e-folding energy and fixes one power-law index at +2. Both models fit the continuum well, but for brevity, we present the results from the \texttt{FDCut} model in this work.

Our final model to fit all observations is: “\texttt{TBabs} $\times$ (\texttt{powerlaw} $\times$ \texttt{FDCut} + \texttt{gauss} + \texttt{bbody}) $\times$ \texttt{gabs}$_{\mathrm{cyc}}$”. The fits are shown in Fig. \ref{fig:unfolded-avg-spectra-residues}. The unfolded spectrum with the best fit of \texttt{FDCut} model is shown on the left panel with the spectrum of obs3. The right panels show the residuals. Panel (a) best fit for obs3 in \textbf{SegA} (b) best fit for obs5 in \textbf{SegB}  (c) best fit for obs7 in \textbf{SegC}. In panels (b) and (c), corresponding to obs5 and obs7 from \textbf{SegB} and \textbf{SegC}, we observe a narrow dip around 10 keV, which we attempted to model with a \texttt{gabs} component. For obs5, the $\chi^2$ improved from approximately 813 for 654 d.o.f. to about 738 for 651 d.o.f. The standard \textit{F-test} tool in \textit{XSPEC} is not suitable for evaluating the statistical significance of including a multiplicative component in the continuum model. Instead, it is more appropriate to construct an F-statistic based on the ratio of variances, as discussed in Appendix A of \citet{Orlandini_2012}. To quantify the significance of this feature, we employed the \textit{mpftest} tool\footnote{\label{note1}\url{http://www.physics.wisc.edu/~craigm/idl/down/mpftest.pro}} to calculate the Probability of Chance Improvement (PCI). Our analysis revealed that the PCI exceeded 12\%. The line had a narrow width of around 2 keV and an optical depth of $\sim$0.02 for obs5. For obs7, the fit improvement was even less significant, with the PCI of including the \texttt{gabs} feature at 10 keV being over 31\%. As a result, we did not include this feature in the best-fit model. The spectral parameters of the best fit are shown in Table. \ref{tab:Bestfit-table-spec}. The power law index varies from around 0.87 to around 0.55, with increasing luminosity showing a clear anti-correlation. The value of the cutoff for all the observations lies between 17 and 26 keV, while the blackbody components have a temperature of 0.35 keV on average. We detect the presence of the main cyclotron line around 44 keV in lower luminosity segments, \textbf{SegA} and \textbf{SegC}, and around 39 keV in the higher luminosity segment, \textbf{SegB} showing a clear negative correlation with the luminosity.

\begin{figure*}
    \centering
	\includegraphics[scale=0.3, trim={0 3.0cm 0 1.8cm},  angle=-90]{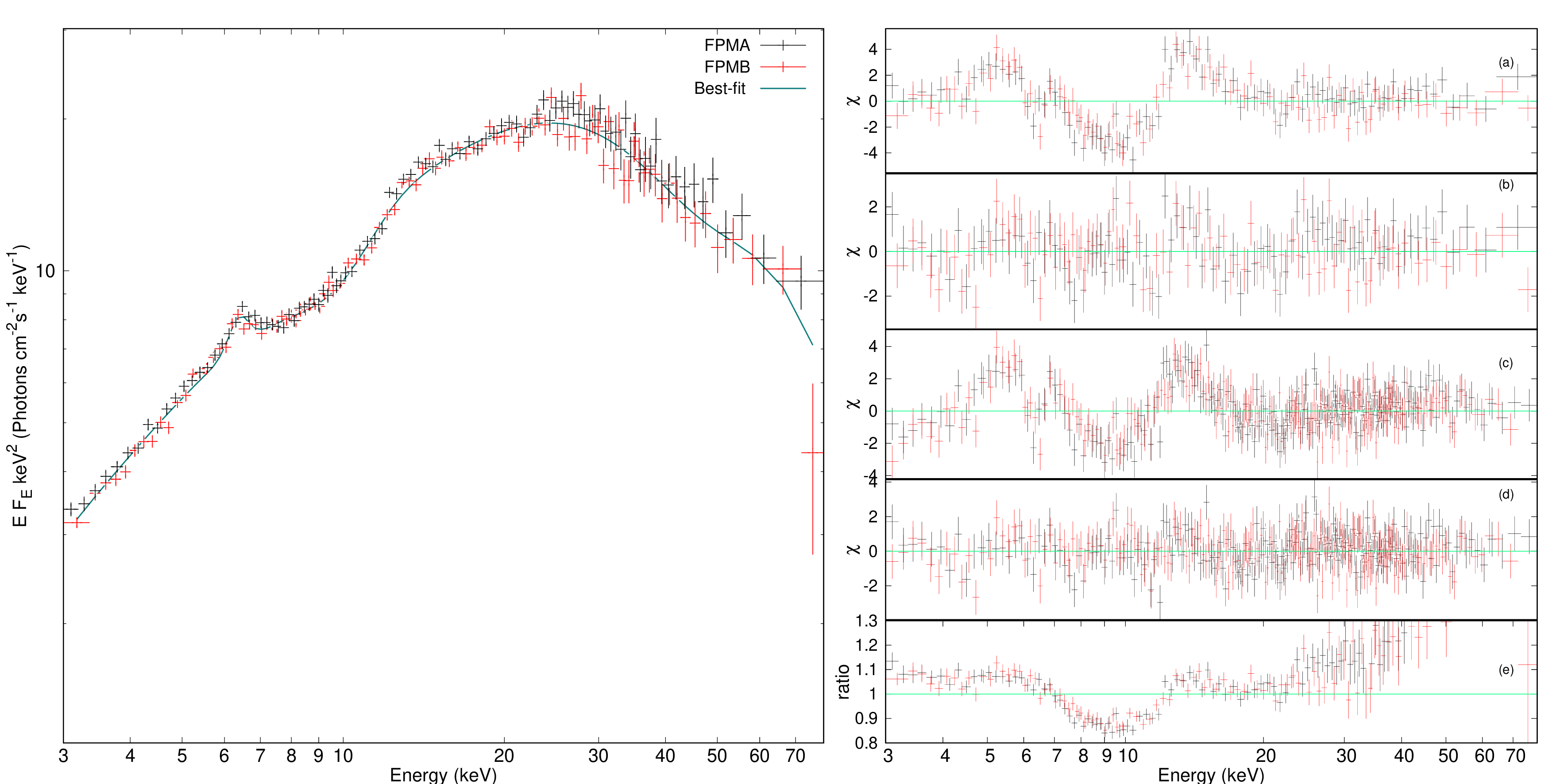}
    \caption{ Phase-resolved spectra: The unfolded spectrum with the best-fit of \texttt{FDCut} model is shown on the left panel with the spectrum of the second phase bin of obs3. The right panel shows the residuals. Panel (a) residuals of the fit with only the phase-average \texttt{FDCut} model (b) best-fit after the inclusion of a \texttt{gabs} component around 9 keV with \texttt{FDCut} model (c) residuals of the fit with only the phase-average \texttt{NPEX} model (d) the best fit after the inclusion of a \texttt{gabs} component around 9 keV with \texttt{NPEX} model (e) the ratio of the data in this phase bin to the best-fit phase-average model which does not have an absorption component around 9 keV. The red and black points correspond to data from FPMA and FPMB, respectively.}
    \label{fig:unfolded-prs-obs3-spectra-residues} 
\end{figure*}

\subsection{Phase-resolved Spectroscopy}
The phase-resolved spectroscopy of this source, focusing particularly on the 44 keV line and its harmonic, was performed by \citet{Kong_2022} using \hxmt \ observations. They have detected the line significantly in all the pulse phases, with two out of ten phase bins showing a weaker fundamental with a stronger harmonic component. Here, we focus on the detection and the pulse phase dependence of a new transient line that appears at lower energies. 

We extracted spectra in 30-phase bins for each observation using the pulse periods of the respective segments mentioned in the previous section. In the 0.16 pulse phase range, centered at the 0.2 phase dip present for all the observations, we detected the presence of a new cyclotron line. In all phase-resolved fits, we fixed the $N_H$, blackbody temperature, and the centroid and width of the iron line to their phase-average values while allowing the normalization parameters to vary. In some of the phase bins that had low statistics at higher energies, we froze the 44 keV line's centroid to the respective phase average values. As a typical case, we describe here the spectrum of the second phase bin in the detected range of obs3 as shown in Fig. \ref{fig:unfolded-prs-obs3-spectra-residues}.  Fitting the data with the phase-average best-fit model alone resulted in clear residuals around 9 keV for this phase bin, with an unfavorable $\chi^2 \sim 1044.99$ for 397 d.o.f., as seen in panel (a) of Fig. \ref{fig:unfolded-prs-obs3-spectra-residues}. Adding a \texttt{gabs} component at this energy significantly improved the fit, reducing the $\chi^2$ to around 396 for 394 d.o.f., as shown in panel (b). To further explore the impact of the continuum chosen, we also fit the data with the \texttt{NPEX} model, which produced wavy residuals (panel c). Including a \texttt{gabs} component in this model improved the residuals, as seen in panel (d). Finally, panel (e) shows the ratio of the data in this phase bin to the phase-average best-fit model, illustrating the transient nature of this cyclotron line, as such a feature is not significant in the phase-average spectrum.

We detected the presence of this cyclotron line only in a narrow set of phase bins covering a 0.16 pulse phase range around the characteristic dip at the phase 0.2. Using the \textit{mpftest} \footref{note1}, we calculated the PCI for each phase bin, reporting the line only when the PCI was below 5\%. For the above-mentioned case, the PCI was approximately $10^{-21}$. The panels below the pulse profiles in Fig. \ref{fig:transient-line-variation} show how this transient line varies across the pulse phase. The line was detected in 5 phase bins in \textbf{SegA} and \textbf{SegB}, and in 4 bins in \textbf{SegC}. Its energy ranged from around 12 keV to 5 keV in \textbf{SegA}, 17 keV to 9 keV in \textbf{SegB}, and 15 keV to 9 keV in \textbf{SegC}. Both the width and optical depth of the feature showed significant variation with pulse phase, with the width ranging from $\sim$0.9 keV to 2.3 keV and the optical depth increasing from 0.1, peaking at 0.21 and then decreasing back to 0.1. No such feature was detected in any of the other phase bins, which were fit well by the phase-average model alone.

\begin{figure*}
    \centering
	\includegraphics[scale=0.3, trim={0 3.0cm 0 -1cm},  angle=-90]{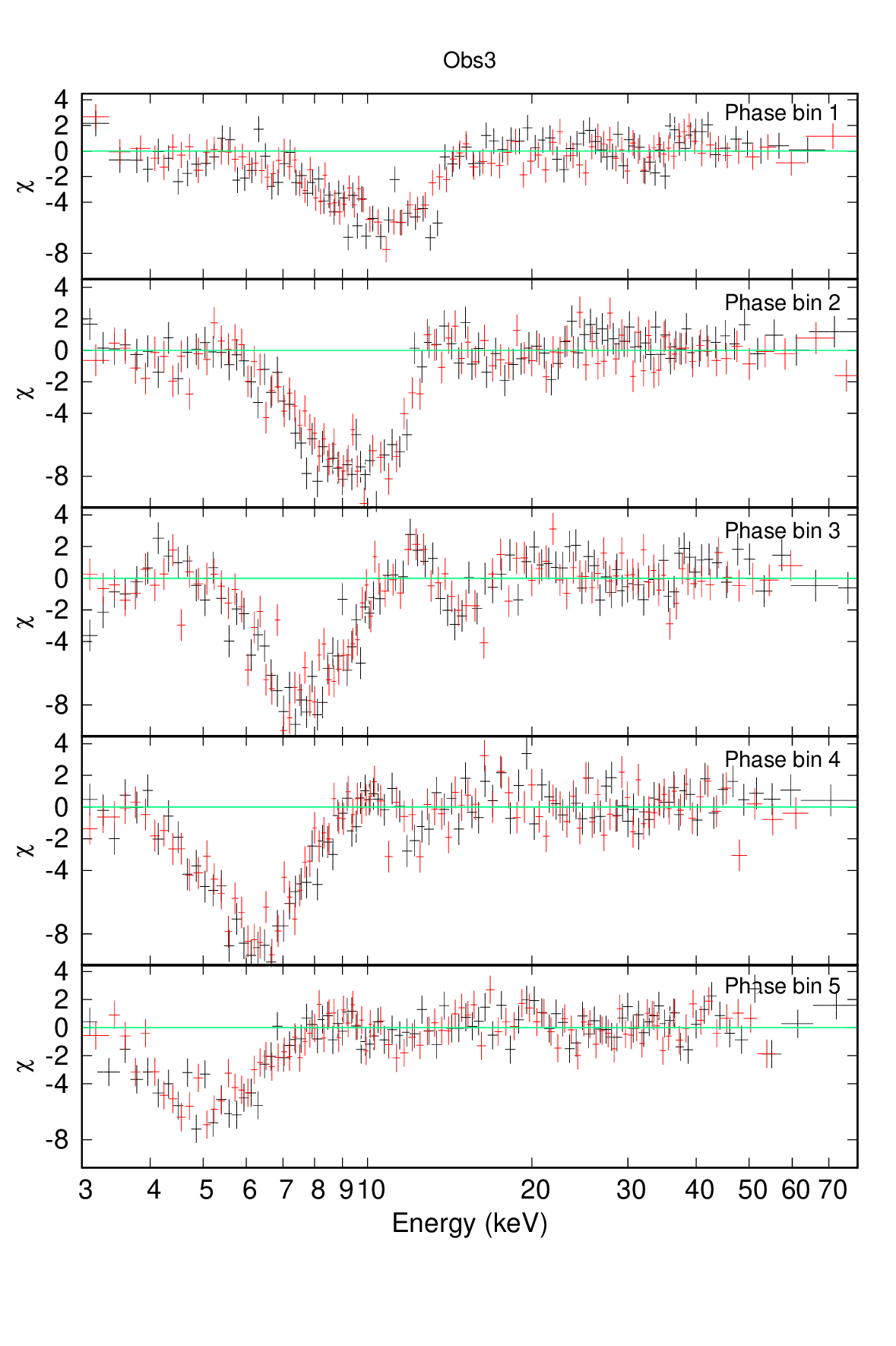}
 	\includegraphics[scale=0.3, trim={0 3.0cm 0 -1cm},  angle=-90]{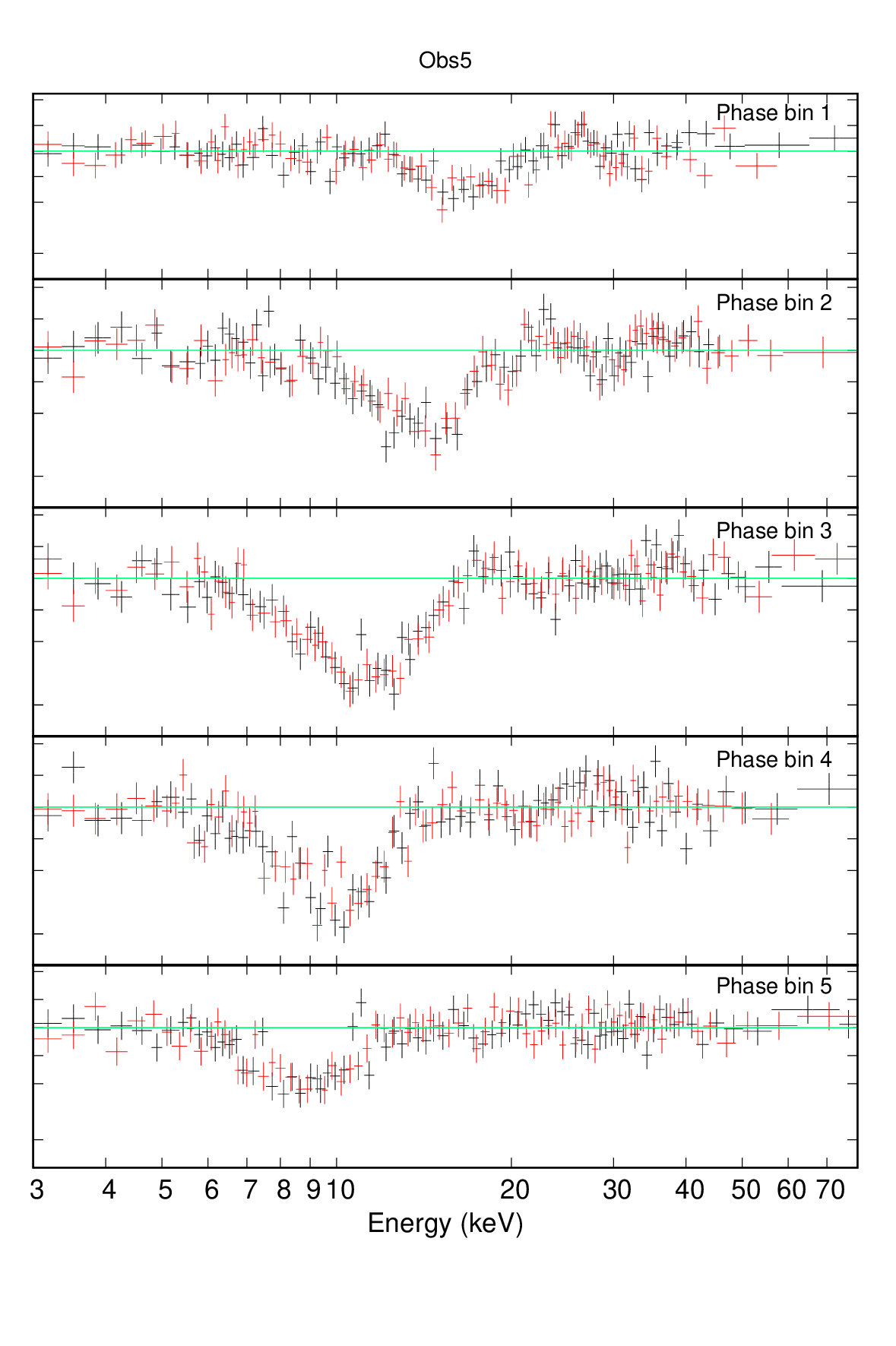}
   \includegraphics[scale=0.3, trim={0 3.0cm 0 -1cm},  angle=-90]{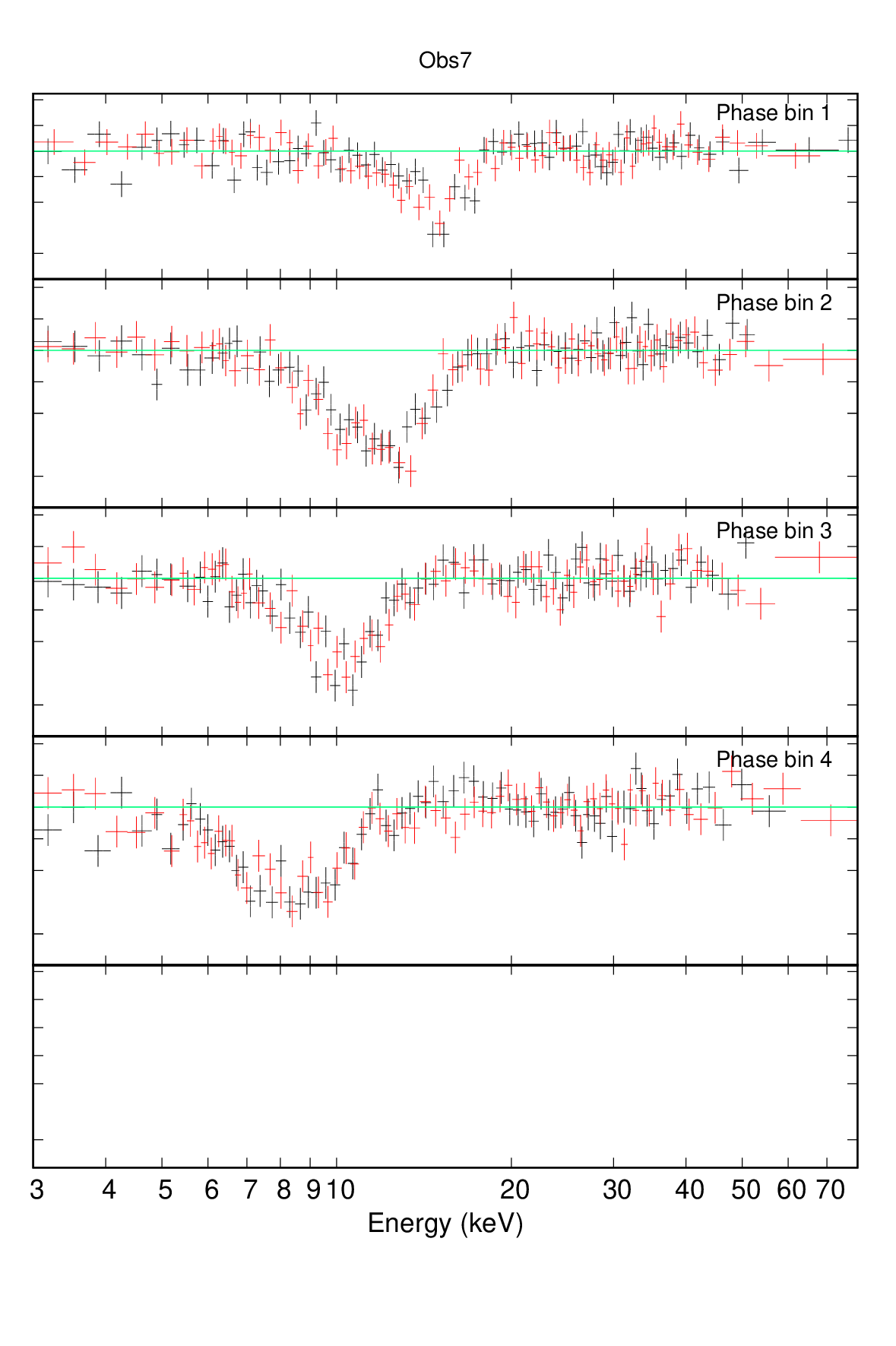}
    \caption{ Phase-resolved spectral residuals: The residuals for each phase bin of obs3, obs5 and obs7 where the transient cyclotron line has been detected, setting the optical depth of the \texttt{gabs} to 0 in the best-fit model. The shift of the line's centroid to lower energies can be clearly seen here. The red and black points correspond to data from FPMA and FPMB, respectively.}
    \label{fig:transient-line-shift-residuals} 
\end{figure*}

\begin{figure*}
    \centering
	\includegraphics[scale=0.6, trim={0 0cm 0 0cm}]{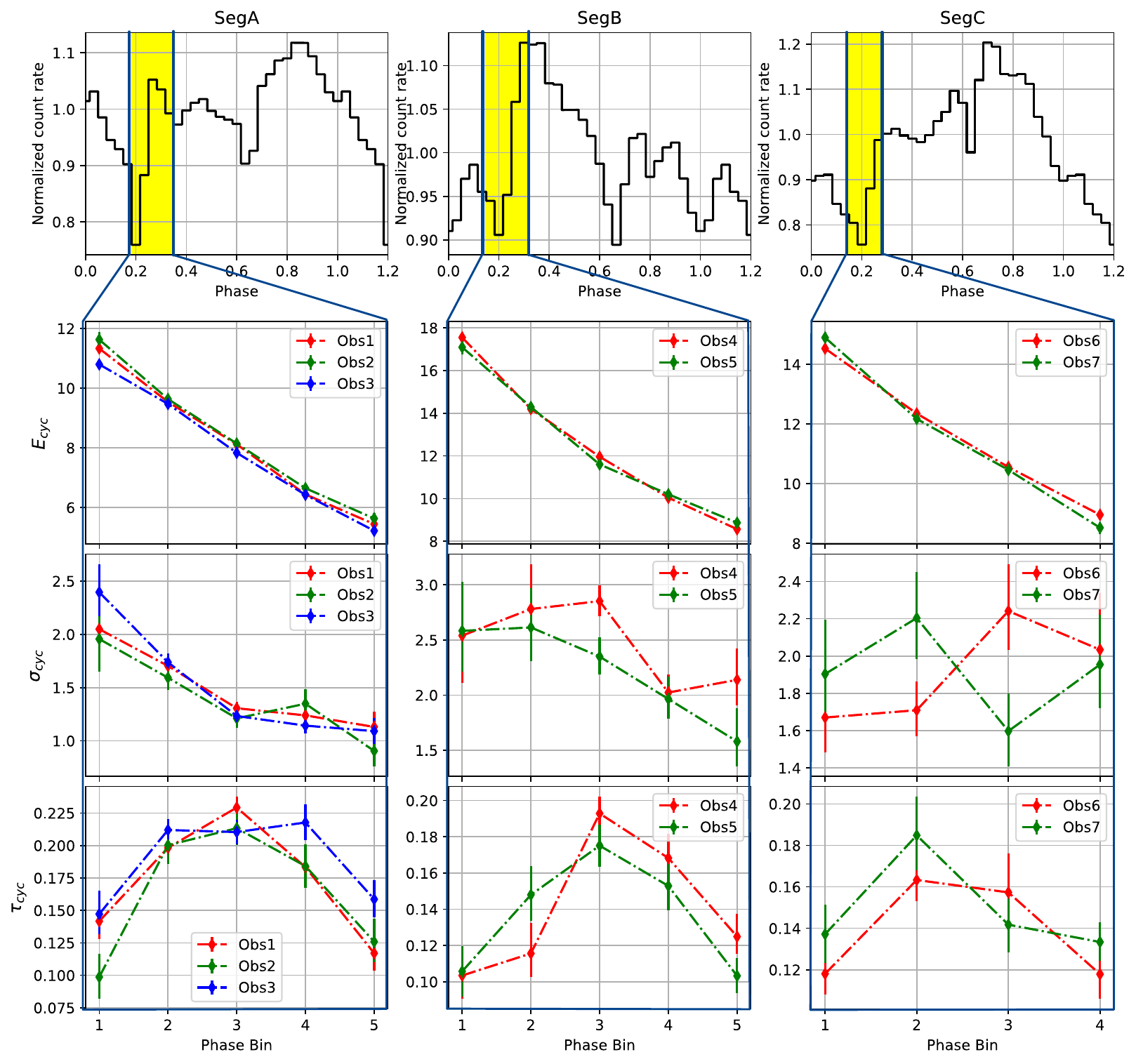}
    \caption{The top 3 panels show the pulse profiles of \textbf{SegA}, \textbf{SegB} and \textbf{SegC} in 30 phase bins. The transient cyclotron line was detected in the highlighted phase bins. Each of the plots below the pulse profiles show the variation of the centroid line energy, the width and the optical depth with pulse phase for the observations in each segment. 1 $\sigma$ uncertainties have been shown on each parameter value. The line was detected in 5 phase bins in \textbf{SegA} and  \textbf{SegB} and in 4 phase bins in \textbf{SegC}.}
    \label{fig:transient-line-variation} 
\end{figure*}

\section{Discussion and conclusion} \label{sec:discussion_conclusion}

\begin{figure*}
    \centering
 \vspace{0.5cm}
	\includegraphics[angle=90,scale=0.58,trim={1.6cm 6.0cm 0cm 0 2.0cm},  angle=-90]{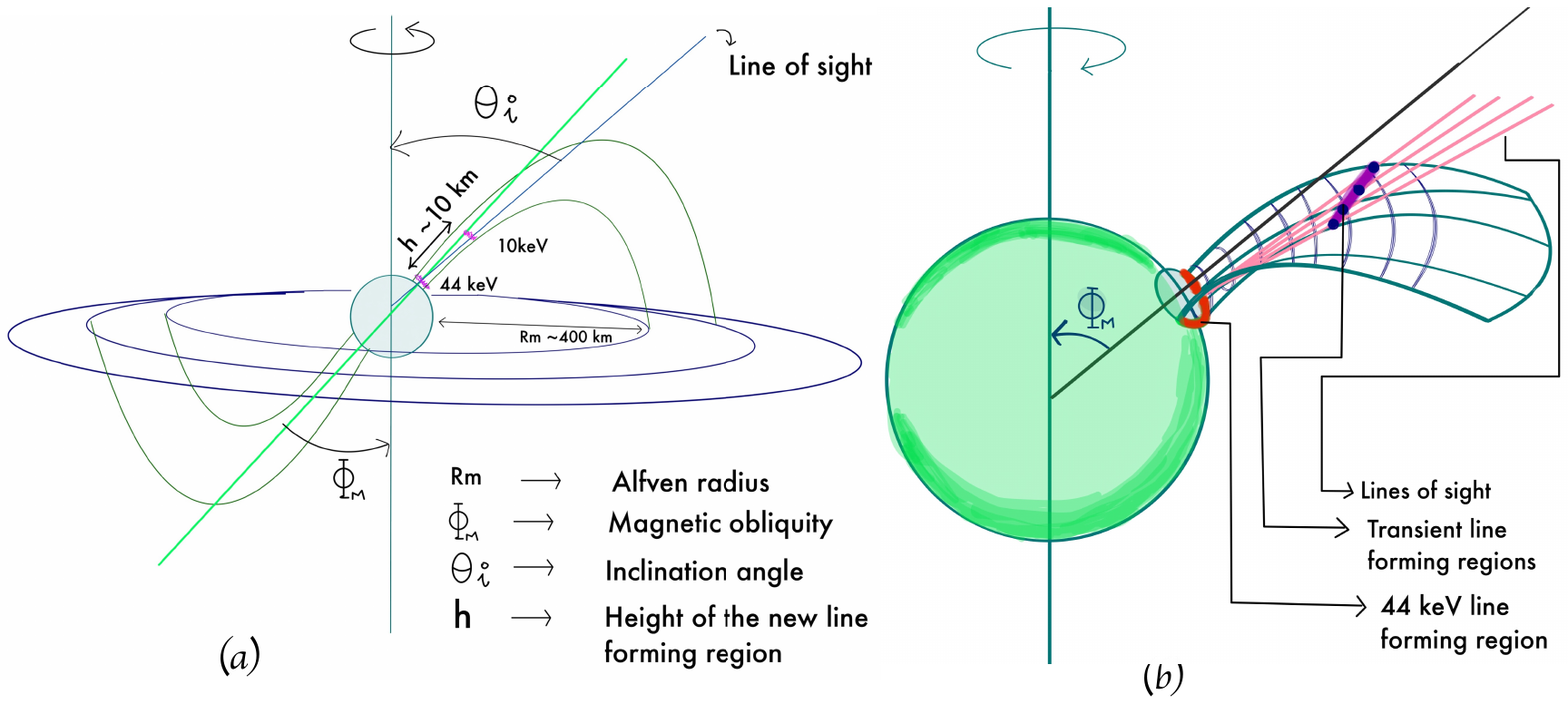}
    \caption{Accretion geometry of A0535+26 (not to scale). Left: Panel (a) shows the approximate Alfven radius and the height of the new cyclotron line forming region. Right: Panel (b) shows a closer view of the NS where the pink lines indicate the different lines of sight passing through two cyclotron line-forming regions, the main CRSF region near the base and the transient lines formed several kilometers above the surface when the NS rotates.}
    \label{fig:geometry-system}
\end{figure*}

Phase-transient cyclotron lines are CRSFs that have been detected in only a narrow range of pulse phases but seldom show their presence in the phase average spectra. Such features have been previously reported for only two accreting pulsars, namely, GRO J2058+42 and SWIFT J1626.6-5156. For GRO J2058+42, \citet{Kabiraj_2020} estimated a magnetic field of $3.7 \times 10^{13}$ G using the accretion torque model \citep{Ghosh_1978}, as the phase-average spectrum showed no cyclotron line. However, through phase-resolved spectroscopy, \citet{Molkov_2019} detected a 10 keV line and two harmonics at 20 and 30 keV in only 10\% of phases, attributing this to the partial visibility of the accretion column. In SWIFT J1626.6-515, \citet{DeCesar_2013} detected a transient 18 keV line in 25\% of the phases, interpreting it as the first harmonic of a 10 keV line seen in the phase-average spectrum. Later, \citet{Molkov_2021}, using \nustar\ data, reported a fundamental line at 5 keV and three harmonics in the phase-average spectrum. 

In the case of A0535+26, although there are weak residuals around 10 keV in the phase-average spectrum of \textbf{SegB} and \textbf{SegC} and the first occurrence of the transient line is around 10 keV in phase-resolved spectra of all the observations of \textbf{SegA}, it is unlikely that this is the standard `10 keV feature' as seen in sources like XTE J1946+274 \citep{Devaraj_2022b,devaraj_2023}. The `10 keV feature'  feature is an absorption-like feature around 10 keV that was found to be independent of time, pulse phase, flux, and the instrument with which it was observed \citep{Coburn_2002,hemanth_2023}. The general consensus for the presence of this feature around 10 keV in several X-ray pulsars has been attributed to the inadequacies of applying the simple phenomenological continuum models to describe the high-quality spectra from sensitive modern instruments. We would like to point out that the transient line shows significant variation with pulse phase and luminosity, which is atypical of the `10 keV' feature. 

Cyclotron lines provide insight into the magnetic field strength in the regions where they form, and phase-resolved spectroscopy provides deep insights into the accretion geometry of the system. We summarize the key results of this work as follows:
\begin{itemize}
    \item A new phase-transient absorption line was detected in all the \nustar \ observations of the source around the characteristic minima of the pulse profile and was detected in only 16 \% of the pulse phases. 
    \item The transient cyclotron line's energy varies significantly as a function of the pulse phase. In \textbf{SegA}, it moves from around 12 keV to 5 keV; from about 17 keV to around 9 keV in the \textbf{SegB}, and from around 15 keV to 9 keV in \textbf{SegC}, indicating a 40 - 58 \% variation.
    \item In addition to pulse phase variations, the transient lines show a clear dependence on luminosity. The highest detected line energy in \textbf{SegA} is around 12 keV, while in \textbf{SegB}, which corresponds to a higher luminosity, it is approximately 17 keV, demonstrating a positive correlation. Overall, the range of line energies shows a consistent positive correlation with luminosity. 
\end{itemize}

In A0535+26, the low-energy line has a large variation in energy with pulse phase and hence, it is unlikely that the 44 keV line is a harmonic of the low-energy line like in the case of SWIFT J1626.6-515. \citet{Kong_2022} detailed the phase dependence of the 44 keV line and its 90 keV harmonic. The detection of low-energy lines in a narrow 0.16 phase range around the characteristic dip at phase 0.2, in addition to the well-studied 44 keV line, suggests the presence of multiple cyclotron line-forming regions similar to GX 301-2 \citep{Furst_2018}. The highest measured value of the line in A0535+26 is $\sim 52$ keV, indicating that the surface magnetic field strength, $B_* \gtrsim$  5 $\times 10^{12}$G \citep{Caballero_2008}. The Alfven radii where the inner region of the accretion disk terminates for the range of luminosities observed with \nustar \ is between 330 - 540 km for canonical NS parameters assuming a $B_* =$  5 $\times 10^{12}$G (using eqn. 18 from \citealt{Becker_2012}). 

A negative correlation of the \ecyc with \Lx was observed thus putting it in the radiation shock regime \citep{Kong_2021, Mandal_2022}. Using eqn. (16) from \citet{Becker_2012},
\begin{equation}
    H = 1.14 \times 10^5 \, \text{cm} \left( \frac{M_*}{1.4 \, M_\odot} \right)^{-1} \left( \frac{R_*}{10 \, \text{km}} \right) \left( \frac{L_X}{10^{37} \, \text{erg s}^{-1}} \right)
\end{equation}
where M$_*$ is the mass of the NS, R$_*$ is the radius of the NS, and \Lx \ is the X-ray luminosity, we can estimate the height $H$ of the radiation shock at the peak of the outburst to be around 14 km above the NS's surface for canonical NS parameters for an \Lx$\sim 1.2 \times 10^{38}$ erg cm$^{-2}$s$^{-1}$ . Assuming a dipolar field configuration with the surface field strength represented by the 52 keV line, the height at which the transient line is formed can be estimated using,
\begin{equation}
    \frac{E_{\text{cyc}}}{E_*} = \left( \frac{R_{\text{*}}}{R_{\text{*}} + h} \right)^3
\end{equation}
where E$_*$ is the energy of the line at the NS's surface, $h$ is the height above the NS where the E$_{\mathrm{cyc}}$ line is formed. The range of heights at which the transient lines between 17 keV and 9 keV are formed lie between 4.5 km and 8 km with the main line (44 keV) forming at 0.6 km, indicating that the transient lines are formed within the accretion column itself below the height at which the radiation shock is formed. The height at which the CRSF forms in the supercritical regime is expected to be significantly closer to the neutron star surface than the altitude of the radiation-dominated shock, as indicated by Eq. 41 of \citet{Becker_2012}. The observed 44 keV CRSF exhibits a negative correlation with luminosity, suggesting it originates in a region near the stellar surface where photons are trapped by advection and ultimately escape through the column walls, producing a fan beam emission. In contrast, the low-energy 10 keV CRSF shows a positive correlation with luminosity, indicating it forms in a different region, possibly at higher altitudes influenced by the density profile above the shock.

The beam pattern in this source is, therefore, likely complex, dominated by a fan beam consistent with the supercritical regime, as described by \citet{Becker_2012}. This explains the detection of the 44 keV CRSF across all phase bins. However, the appearance of the transient 10 keV CRSF only near the pulse profile's minima suggests the presence of a subdominant pencil beam component, where a fraction of the emission is directed along the column axis and absorbed at higher altitudes. This configuration points to the possibility that transient CRSFs form in regions distinct from the traditional shock-formation zones. The precise locations and behavior of these features, along with their luminosity dependence, likely depend on the post-shock fluid dynamics and require detailed analytical modeling to fully understand the interplay between these emission components and the CRSF formation regions.

To explain the findings of this work while accounting for the previous results involving A0535+26, we suggest a geometry where the accretion column sweeps the line of sight (see Fig. \ref{fig:geometry-system}). The main cyclotron line is observed at all phases, whereas the transient line is only observed in the 0.16 pulse phase range, centered at the characteristic dip at the phase 0.2. We show a diagram of the system in Fig. \ref{fig:geometry-system}, panel (a), where $\Phi_M$ is the magnetic obliquity, $\theta_i$ is the inclination angle of the system, and the magnetospheric radius is around 400 km on average. At the same time, the 44 keV line is formed close to the surface, while the 9 keV transient line forms approximately 10 km above the surface of the NS. The simultaneous observation of the 44 keV line and the transient line occurs during the phase when the accretion column crosses the line of sight.

The drastic variation of the transient line’s energy with the pulse phase can be explained if our line of sight cuts across different heights of the accretion column, as shown in Fig. \ref{fig:geometry-system}, panel (b). Additionally, the variation of the transient line’s energy with luminosity is likely due to changes in the Alfvén radius, which alters the arc of the accretion stream. At lower luminosities, the Alfvén radius is further away from the NS, resulting in a larger arc and a higher line-forming region. Conversely, at higher luminosities, an increased accretion rate causes the Alfvén radius to shrink, leading to a tighter arc and a line-forming region closer to the NS.

Previous results of phase-resolved spectroscopy using \textit{Suzaku} data show a dramatic increase in the column density of the partial covering model at the characteristic dip (see Fig. 7 of \citealt{Maitra2013}), an effect not observed at other pulse phases. The other dips in the pulse profiles are likely caused by the evolution of the emission geometries and beaming patterns, while the characteristic dip is due to the absorption of X-rays by the accretion column itself. The inclination angle of this system was determined to be around 37$^\circ$ \citep{Giovannelli_2007}. Assuming the spin axis of the NS is perpendicular to the orbital plane and the accretion column indeed sweeps the line of sight, the magnetic obliquity is likely close to or slightly larger than this value. 

\newpage
\section*{Acknowledgement}
\begin{acknowledgments}
We are grateful to the referee for their constructive comments, which improved the clarity and presentation of this work. This research utilized the \textit{NuSTAR} data analysis software (\texttt{NuSTARDAS}), a joint development by the ASI Science Data Center (ASDC, Italy) and the California Institute of Technology (USA). We also acknowledge the use of data and software provided by the High Energy Astrophysics Science Archive Research Center (HEASARC), a service of NASA’s Astrophysics Science Division at the Goddard Space Flight Center (GSFC).
\end{acknowledgments}


\begin{turnpage}  
\begin{table}[htb]
\bgroup
\caption{Best-fitting phase-averaged spectral parameters using the FDcut model. Errors are reported at 90\% confidence.}
\begin{threeparttable}
\renewcommand{\arraystretch}{1.8}  
\begin{tabular}{|l|l|l|l|l|l|l|l|}
\hline
\hline
Parameters                                                & Obs1                             & Obs2                              & Obs3                              & Obs4                              & Obs5                              & Obs6                              & Obs7                              \\
\hline
\hline
C$_{B}$                                                  & $0.9823_{-0.0015}^{+0.0015}$             & $0.9801_{-0.002}^{+0.002}$                & $0.9749_{-0.0014}^{+0.0014}$              & $0.9769_{-0.0012}^{+0.0012}$              & $0.9788_{-0.0016}^{+0.0016}$              & $0.9664_{-0.0012}^{+0.0012}$              & $0.9684_{-0.0016}^{+0.0016}$              \\
N$_\mathrm{H}$[$\times 10^{22}$cm$^{-2}$]                & $1.26_{-0.39}^{+0.44}$                   & $2.07_{-0.48}^{+0.47}$                    & $1.37_{-0.46}^{+0.51}$                    & $0.301_{-0.3}^{+0.47}$                    & $0.928_{-0.6}^{+0.58}$                    & $1.05_{-0.32}^{+0.35}$                    & $0.406_{-0.34}^{+0.33}$                   \\
PowLaw Index $\Gamma$                                  & $0.843_{-0.029}^{+0.029}$                & $0.8797_{-0.038}^{+0.037}$                & $0.8157_{-0.03}^{+0.03}$                  & $0.5549_{-0.018}^{+0.02}$                 & $0.573_{-0.031}^{+0.028}$                 & $0.6397_{-0.023}^{+0.023}$                & $0.6269_{-0.033}^{+0.017}$                \\
Powerlaw $\Gamma_{\text{Norm}}$ $^{*}$ & $1.04_{-0.040}^{+0.022}$ & $1.17_{-0.072}^{+0.061}$ & $1.15_{-0.051}^{+0.056}$ & $2.25_{-0.082}^{+0.10}$ & $2.33_{-0.13}^{+0.14}$ & $1.89_{-0.057}^{+0.062}$ & $1.79_{-0.070}^{+0.065}$ \\
FDcut E$_{\mathrm{cut}}$(keV)                                  & $20.8_{-2.7}^{+3.3}$                     & $26.2_{-4.8}^{+9.3}$                      & $19.7_{-2.6}^{+3.2}$                     & $26.4_{-1.9}^{+2.1}$                     & $26.7_{-2.9}^{+3.3}$                     & $17.6_{-1.4}^{+1.6}$                     & $18.4_{-2.4}^{+3.6}$                     \\
FDcut E$_{\mathrm{fold}}$ kT(keV)                              & $14.27_{-0.37}^{+0.31}$                  & $13.52_{-1.7}^{+0.64}$                    & $14.44_{-0.35}^{+0.3}$                   & $10.25_{-0.39}^{+0.33}$                  & $10.23_{-0.63}^{+0.48}$                  & $12.57_{-0.15}^{+0.15}$                  & $12.71_{-0.33}^{+0.25}$                  \\
Blackbody kT (keV)                                       & $0.2944_{-0.074}^{+0.056}$               & $0.35_{-0.069}^{+0.046}$                 & $0.3264_{-0.072}^{+0.049}$               & $0.3601_{-0.11}^{+0.075}$                & $0.4232_{-0.081}^{+0.07}$                & $0.2828_{-0.046}^{+0.04}$                & $0.26_{-0.17}^{+0.09}$                   \\
Blackbody Norm                                           & $0.09874_{-0.057}^{+0.61}$               & $0.07966_{-0.027}^{+0.029}$               & $0.06646_{-0.031}^{+0.17}$               & $0.1093_{-0.056}^{+0.73}$                & $0.1055_{-0.052}^{+0.075}$               & $0.367_{-0.19}^{+0.84}$                  & $0.336_{-0.19}^{+0.21}$                  \\
E$_{\mathrm{Fe}}$   (keV)                                & $6.364_{-0.019}^{+0.018}$                & $6.357_{-0.022}^{+0.021}$                 & $6.37_{-0.017}^{+0.016}$                 & $6.457_{-0.014}^{+0.015}$                & $6.448_{-0.022}^{+0.022}$                & $6.469_{-0.02}^{+0.02}$                  & $6.426_{-0.028}^{+0.027}$                \\
$\sigma_{\mathrm{Fe}}$  (keV)                                  & $0.1434_{-0.047}^{+0.045}$               & $0.1173_{-0.063}^{+0.057}$                & $0.1778_{-0.045}^{+0.043}$               & $0.329_{-0.022}^{+0.023}$                & $0.3307_{-0.035}^{+0.036}$               & $0.2909_{-0.029}^{+0.03}$                & $0.3301_{-0.053}^{+0.054}$               \\
Norm$_{\mathrm{Fe}}$[$\times 10^{-3}$]$^{\dagger}$                   & $7.334_{-0.9}^{+0.99}$                   & $8.042_{-0.0011}^{+0.0014}$               & $9.529_{-0.0012}^{+0.0013}$              & $70.26_{-0.0048}^{+0.0049}$              & $66.47_{-0.0068}^{+0.0078}$              & $26.05_{-0.0024}^{+0.0025}$              & $30.57_{-0.0039}^{+0.0042}$              \\
Fe equivalent width [eV] & $46.0_{-5.6}^{+5.4}$ & $43.0_{-5.3}^{+6.6}$ & $53.0_{-6.8}^{+6.2}$ & $100.0_{-6.1}^{+10.0}$ & $95.0_{-9.0}^{+16.0}$ & $64.0_{-7.1}^{+6.6}$ & $76.0_{-8.0}^{+8.4}$ \\
E$_{\texttt{cyc}}$   (keV)                               & $43.64_{-0.78}^{+0.84}$                  & $44.22_{-1.05}^{+1.12}$                   & $44.66_{-0.75}^{+0.8}$                  & $38.95_{-0.52}^{+0.54}$                 & $39.3_{-0.74}^{+0.8}$                   & $43.48_{-0.84}^{+0.9}$                  & $45.1_{-1.2}^{+1.4}$                    \\
$\sigma_{\texttt{cyc}}$ (keV)                                 & $11.3_{-0.92}^{+1.2}$                    & $13.1_{-1.6}^{+2.2}$                      & $12.05_{-0.92}^{+1.1}$                  & $13.07_{-0.79}^{+0.74}$                 & $13.6_{-1.1}^{+1.1}$                    & $10.87_{-0.91}^{+1.1}$                  & $12.7_{-1.5}^{+2.1}$                    \\
$\tau_{\texttt{cyc}}$                                    & $0.4232_{-0.046}^{+0.066}$               & $0.505_{-0.1}^{+0.29}$                    & $0.4205_{-0.046}^{+0.063}$              & $0.4675_{-0.078}^{+0.098}$              & $0.483_{-0.11}^{+0.16}$                 & $0.2605_{-0.029}^{+0.038}$              & $0.3146_{-0.054}^{+0.098}$              \\
\hline
Tot-Chi-sq(d.o.f)                                & $857(681)$                                    & $680(640)$                                     & $806(693)$                                    & $1012(702)$                                   & $813(654)$                                    & $958(693)$                                    & $783(658)$                                   \\
Reduced Chi-Square                               & $1.26$                                       & $1.06$                                        & $1.16$                                        & $1.44$                                        & $1.24$                                        & $1.38$                                        & $1.19$                                        \\
\hline
Flux [\(\times 10^{-8}\) erg cm$^{-2}$ s$^{-1}$]$^\ddagger$ & $4.95_{-0.011}^{+0.011}$ & $5.63_{-0.016}^{+0.016}$ & $5.67_{-0.011}^{+0.011}$ & $26.2_{-0.040}^{+0.040}$ & $25.6_{-0.054}^{+0.055}$ & $14.1_{-0.024}^{+0.024}$ & $14.2_{-0.032}^{+0.032}$ \\
L$_X$ [ $\times$ 10$^{37}$ ergs s$^{-1}$ ] &    2.37    & 2.70  &    2.71    &       12.54      &       12.25       &       6.75        &       6.80    \\
\hline
\end{tabular}  
\begin{tablenotes}
 \item[] $^*$ In units of photons keV$^{-1}$ cm$^{-2}$ s$^{-1}$
 \item[] $^\dagger$ In units of photons cm$^{-2}$ s$^{-1}$ 
 \item[] $^\ddagger$ Flux in 3-79 keV band 
\end{tablenotes}
\end{threeparttable}
\egroup
\label{tab:Bestfit-table-spec}
\end{table}
\end{turnpage}

\vspace{5mm}






\newpage
\bibliography{bibtex}{}

\begin{thebibliography}{}
\expandafter\ifx\csname natexlab\endcsname\relax\def\natexlab#1{#1}\fi
\providecommand{\url}[1]{\href{#1}{#1}}
\providecommand{\dodoi}[1]{doi:~\href{http://doi.org/#1}{\nolinkurl{#1}}}
\providecommand{\doeprint}[1]{\href{http://ascl.net/#1}{\nolinkurl{http://ascl.net/#1}}}
\providecommand{\doarXiv}[1]{\href{https://arxiv.org/abs/#1}{\nolinkurl{https://arxiv.org/abs/#1}}}

\bibitem[{{Becker} {et~al.}(2012){Becker}, {Klochkov}, {Sch{\"o}nherr}, {Nishimura}, {Ferrigno}, {Caballero}, {Kretschmar}, {Wolff}, {Wilms}, \& {Staubert}}]{Becker_2012}
{Becker}, P.~A., {Klochkov}, D., {Sch{\"o}nherr}, G., {et~al.} 2012, \aap, 544, A123, \dodoi{10.1051/0004-6361/201219065}

\bibitem[{{Caballero} {et~al.}(2007){Caballero}, {Kretschmar}, {Santangelo}, {Staubert}, {Klochkov}, {Camero}, {Ferrigno}, {Finger}, {Kreykenbohm}, {McBride}, {Pottschmidt}, {Rothschild}, {Sch{\"o}nherr}, {Segreto}, {Suchy}, {Wilms}, \& {Wilson}}]{Caballero_2007}
{Caballero}, I., {Kretschmar}, P., {Santangelo}, A., {et~al.} 2007, \aap, 465, L21, \dodoi{10.1051/0004-6361:20067032}

\bibitem[{{Caballero} {et~al.}(2008){Caballero}, {Santangelo}, {Kretschmar}, {Staubert}, {Postnov}, {Klochkov}, {Camero-Arranz}, {Finger}, {Kreykenbohm}, {Pottschmidt}, {Rothschild}, {Suchy}, {Wilms}, \& {Wilson}}]{Caballero_2008}
{Caballero}, I., {Santangelo}, A., {Kretschmar}, P., {et~al.} 2008, \aap, 480, L17, \dodoi{10.1051/0004-6361:20079310}

\bibitem[{{Chhotaray} {et~al.}(2023){Chhotaray}, {Jaisawal}, {Kumari}, {Naik}, {Kumar}, \& {Jana}}]{Chhotaray_2023}
{Chhotaray}, B., {Jaisawal}, G.~K., {Kumari}, N., {et~al.} 2023, \mnras, 518, 5089, \dodoi{10.1093/mnras/stac3354}

\bibitem[{{Coburn} {et~al.}(2002){Coburn}, {Heindl}, {Rothschild}, {Gruber}, {Kreykenbohm}, {Wilms}, {Kretschmar}, \& {Staubert}}]{Coburn_2002}
{Coburn}, W., {Heindl}, W.~A., {Rothschild}, R.~E., {et~al.} 2002, \apj, 580, 394, \dodoi{10.1086/343033}

\bibitem[{{DeCesar} {et~al.}(2013){DeCesar}, {Boyd}, {Pottschmidt}, {Wilms}, {Suchy}, \& {Miller}}]{DeCesar_2013}
{DeCesar}, M.~E., {Boyd}, P.~T., {Pottschmidt}, K., {et~al.} 2013, \apj, 762, 61, \dodoi{10.1088/0004-637X/762/1/61}

\bibitem[{{Devaraj} \& {Paul}(2022)}]{Devaraj_2022b}
{Devaraj}, A., \& {Paul}, B. 2022, \mnras, 517, 2599, \dodoi{10.1093/mnras/stac2806}

\bibitem[{Devaraj {et~al.}(2023)Devaraj, Sharma, Nagesh, \& Paul}]{devaraj_2023}
Devaraj, A., Sharma, R., Nagesh, S., \& Paul, B. 2023, Monthly Notices of the Royal Astronomical Society, 527, 11015, \dodoi{10.1093/mnras/stad3886}

\bibitem[{{F{\"u}rst} {et~al.}(2018){F{\"u}rst}, {Falkner}, {Marcu-Cheatham}, {Grefenstette}, {Tomsick}, {Pottschmidt}, {Walton}, {Natalucci}, \& {Kretschmar}}]{Furst_2018}
{F{\"u}rst}, F., {Falkner}, S., {Marcu-Cheatham}, D., {et~al.} 2018, \aap, 620, A153, \dodoi{10.1051/0004-6361/201732132}

\bibitem[{{Ghosh} \& {Lamb}(1978)}]{Ghosh_1978}
{Ghosh}, P., \& {Lamb}, F.~K. 1978, \apjl, 223, L83, \dodoi{10.1086/182734}

\bibitem[{{Giovannelli} {et~al.}(2007){Giovannelli}, {Bernabei}, {Rossi}, \& {Sabau-Graziati}}]{Giovannelli_2007}
{Giovannelli}, F., {Bernabei}, S., {Rossi}, C., \& {Sabau-Graziati}, L. 2007, \aap, 475, 651, \dodoi{10.1051/0004-6361:20066149}

\bibitem[{Harrison {et~al.}(2013)Harrison, Craig, Christensen, Hailey, Zhang, Boggs, Stern, Cook, Forster, Giommi, Grefenstette, Kim, Kitaguchi, Koglin, Madsen, Mao, Miyasaka, Mori, Perri, Pivovaroff, Puccetti, Rana, Westergaard, Willis, Zoglauer, An, Bachetti, Barri{\`{e}}re, Bellm, Bhalerao, Brejnholt, Fuerst, Liebe, Markwardt, Nynka, Vogel, Walton, Wik, Alexander, Cominsky, Hornschemeier, Hornstrup, Kaspi, Madejski, Matt, Molendi, Smith, Tomsick, Ajello, Ballantyne, Balokovi{\'{c}}, Barret, Bauer, Blandford, Brandt, Brenneman, Chiang, Chakrabarty, Chenevez, Comastri, Dufour, Elvis, Fabian, Farrah, Fryer, Gotthelf, Grindlay, Helfand, Krivonos, Meier, Miller, Natalucci, Ogle, Ofek, Ptak, Reynolds, Rigby, Tagliaferri, Thorsett, Treister, \& Urry}]{Harrison_2013}
Harrison, F.~A., Craig, W.~W., Christensen, F.~E., {et~al.} 2013, The Astrophysical Journal, 770, 103, \dodoi{10.1088/0004-637x/770/2/103}

\bibitem[{{Hu} {et~al.}(2023){Hu}, {Ji}, {Yu}, {Wang}, {Doroshenko}, {Santangelo}, {Saathoff}, {Zhang}, {Zhang}, \& {Kong}}]{Hu_2023}
{Hu}, Y.~F., {Ji}, L., {Yu}, C., {et~al.} 2023, \apj, 945, 138, \dodoi{10.3847/1538-4357/acbc7a}

\bibitem[{{Kaastra} \& {Bleeker}(2016)}]{Kaastra_2016}
{Kaastra}, J.~S., \& {Bleeker}, J.~A.~M. 2016, \aap, 587, A151, \dodoi{10.1051/0004-6361/201527395}

\bibitem[{{Kabiraj} \& {Paul}(2020)}]{Kabiraj_2020}
{Kabiraj}, S., \& {Paul}, B. 2020, \mnras, 497, 1059, \dodoi{10.1093/mnras/staa2079}

\bibitem[{{Klochkov} {et~al.}(2011){Klochkov}, {Staubert}, {Santangelo}, {Rothschild}, \& {Ferrigno}}]{Klochkov_2011}
{Klochkov}, D., {Staubert}, R., {Santangelo}, A., {Rothschild}, R.~E., \& {Ferrigno}, C. 2011, \aap, 532, A126, \dodoi{10.1051/0004-6361/201116800}

\bibitem[{{Kong} {et~al.}(2021){Kong}, {Zhang}, {Ji}, {Reig}, {Doroshenko}, {Santangelo}, {Staubert}, {Zhang}, {Soria}, {Chang}, {Chen}, {Wang}, {Tao}, \& {Qu}}]{Kong_2021}
{Kong}, L.~D., {Zhang}, S., {Ji}, L., {et~al.} 2021, \apjl, 917, L38, \dodoi{10.3847/2041-8213/ac1ad3}

\bibitem[{{Kong} {et~al.}(2022){Kong}, {Zhang}, {Ji}, {Doroshenko}, {Santangelo}, {Orlandini}, {Frontera}, {Li}, {Chen}, {Wang}, {Chang}, {Qu}, \& {Zhang}}]{Kong_2022}
{Kong}, L.-D., {Zhang}, S., {Ji}, L., {et~al.} 2022, \apj, 932, 106, \dodoi{10.3847/1538-4357/ac6e66}

\bibitem[{{Maitra}(2017)}]{Maitra_2017}
{Maitra}, C. 2017, Journal of Astrophysics and Astronomy, 38, 50, \dodoi{10.1007/s12036-017-9476-3}

\bibitem[{{Maitra} \& {Paul}(2013)}]{Maitra2013}
{Maitra}, C., \& {Paul}, B. 2013, \apj, 771, 96, \dodoi{10.1088/0004-637X/771/2/96}

\bibitem[{{Mandal} \& {Pal}(2022)}]{Mandal_2022}
{Mandal}, M., \& {Pal}, S. 2022, \mnras, 511, 1121, \dodoi{10.1093/mnras/stac111}

\bibitem[{Manikantan {et~al.}(2023)Manikantan, Paul, \& Rana}]{hemanth_2023}
Manikantan, H., Paul, B., \& Rana, V. 2023, Monthly Notices of the Royal Astronomical Society, 526, 1, \dodoi{10.1093/mnras/stad2527}

\bibitem[{{Molkov} {et~al.}(2021){Molkov}, {Doroshenko}, {Lutovinov}, {Tsygankov}, {Santangelo}, {Mereminskiy}, \& {Semena}}]{Molkov_2021}
{Molkov}, S., {Doroshenko}, V., {Lutovinov}, A., {et~al.} 2021, \apjl, 915, L27, \dodoi{10.3847/2041-8213/ac0c15}

\bibitem[{{Molkov} {et~al.}(2019){Molkov}, {Lutovinov}, {Tsygankov}, {Mereminskiy}, \& {Mushtukov}}]{Molkov_2019}
{Molkov}, S., {Lutovinov}, A., {Tsygankov}, S., {Mereminskiy}, I., \& {Mushtukov}, A. 2019, \apjl, 883, L11, \dodoi{10.3847/2041-8213/ab3e4d}

\bibitem[{{M{\"u}ller} {et~al.}(2013){M{\"u}ller}, {Klochkov}, {Caballero}, \& {Santangelo}}]{Muller_2013}
{M{\"u}ller}, D., {Klochkov}, D., {Caballero}, I., \& {Santangelo}, A. 2013, \aap, 552, A81, \dodoi{10.1051/0004-6361/201220347}

\bibitem[{{Orlandini} {et~al.}(2012){Orlandini}, {Frontera}, {Masetti}, {Sguera}, \& {Sidoli}}]{Orlandini_2012}
{Orlandini}, M., {Frontera}, F., {Masetti}, N., {Sguera}, V., \& {Sidoli}, L. 2012, \apj, 748, 86, \dodoi{10.1088/0004-637X/748/2/86}

\bibitem[{{Priedhorsky} \& {Terrell}(1983)}]{Priedhorsky_1983}
{Priedhorsky}, W.~C., \& {Terrell}, J. 1983, \nat, 303, 681, \dodoi{10.1038/303681a0}

\bibitem[{{Reig}(2011)}]{Reig_2011}
{Reig}, P. 2011, \apss, 332, 1, \dodoi{10.1007/s10509-010-0575-8}

\bibitem[{{Rosenberg} {et~al.}(1975){Rosenberg}, {Eyles}, {Skinner}, \& {Willmore}}]{Rosenberg_1975}
{Rosenberg}, F.~D., {Eyles}, C.~J., {Skinner}, G.~K., \& {Willmore}, A.~P. 1975, \nat, 256, 628, \dodoi{10.1038/256628a0}

\bibitem[{{Rothschild} {et~al.}(2017){Rothschild}, {K{\"u}hnel}, {Pottschmidt}, {Hemphill}, {Postnov}, {Gornostaev}, {Shakura}, {F{\"u}rst}, {Wilms}, {Staubert}, \& {Klochkov}}]{Rothschild_2017}
{Rothschild}, R.~E., {K{\"u}hnel}, M., {Pottschmidt}, K., {et~al.} 2017, \mnras, 466, 2752, \dodoi{10.1093/mnras/stw3222}

\bibitem[{{Sartore} {et~al.}(2015){Sartore}, {Jourdain}, \& {Roques}}]{Sartore_2015}
{Sartore}, N., {Jourdain}, E., \& {Roques}, J.~P. 2015, \apj, 806, 193, \dodoi{10.1088/0004-637X/806/2/193}

\bibitem[{{Sch{\"o}nherr} {et~al.}(2007){Sch{\"o}nherr}, {Wilms}, {Kretschmar}, {Kreykenbohm}, {Santangelo}, {Rothschild}, {Coburn}, \& {Staubert}}]{Schonherr2007}
{Sch{\"o}nherr}, G., {Wilms}, J., {Kretschmar}, P., {et~al.} 2007, \aap, 472, 353, \dodoi{10.1051/0004-6361:20077218}

\bibitem[{{Shui} {et~al.}(2024){Shui}, {Zhang}, {Wang}, {Mushtukov}, {Santangelo}, {Zhang}, {Kong}, {Ji}, {Chen}, {Doroshenko}, {Frontera}, {Chang}, {Peng}, {Yin}, {Qu}, {Tao}, {Ge}, {Li}, {Ye}, \& {Li}}]{Shui_2024}
{Shui}, Q.~C., {Zhang}, S., {Wang}, P.~J., {et~al.} 2024, \mnras, 528, 7320, \dodoi{10.1093/mnras/stae352}

\bibitem[{{Staubert} {et~al.}(2014){Staubert}, {Klochkov}, {Wilms}, {Postnov}, {Shakura}, {Rothschild}, {F{\"u}rst}, \& {Harrison}}]{staubert_2014}
{Staubert}, R., {Klochkov}, D., {Wilms}, J., {et~al.} 2014, \aap, 572, A119, \dodoi{10.1051/0004-6361/201424203}

\bibitem[{{Staubert} {et~al.}(2019){Staubert}, {Tr{\"u}mper}, {Kendziorra}, {Klochkov}, {Postnov}, {Kretschmar}, {Pottschmidt}, {Haberl}, {Rothschild}, {Santangelo}, {Wilms}, {Kreykenbohm}, \& {F{\"u}rst}}]{Staubert_2019}
{Staubert}, R., {Tr{\"u}mper}, J., {Kendziorra}, E., {et~al.} 2019, \aap, 622, A61, \dodoi{10.1051/0004-6361/201834479}

\bibitem[{{Steele} {et~al.}(1998){Steele}, {Negueruela}, {Coe}, \& {Roche}}]{Steele_1998}
{Steele}, I.~A., {Negueruela}, I., {Coe}, M.~J., \& {Roche}, P. 1998, \mnras, 297, L5, \dodoi{10.1046/j.1365-8711.1998.01593.x}

\bibitem[{{Suchy} {et~al.}(2008){Suchy}, {Pottschmidt}, {Wilms}, {Kreykenbohm}, {Sch{\"o}nherr}, {Kretschmar}, {McBride}, {Caballero}, {Rothschild}, \& {Grinberg}}]{Suchy_2008}
{Suchy}, S., {Pottschmidt}, K., {Wilms}, J., {et~al.} 2008, \apj, 675, 1487, \dodoi{10.1086/527042}

\bibitem[{Tanaka(1986)}]{Tanaka_1986}
Tanaka, Y. 1986, in Radiation Hydrodynamics in Stars and Compact Objects, ed. D.~Mihalas \& K.-H.~A. Winkler (Berlin, Heidelberg: Springer Berlin Heidelberg), 198--221.
\newblock \url{http://dx.doi.org/10.1007/3-540-16764-1_12}

\bibitem[{{Tsygankov} {et~al.}(2017){Tsygankov}, {Doroshenko}, {Lutovinov}, {Mushtukov}, \& {Poutanen}}]{Tsygankov_2017}
{Tsygankov}, S.~S., {Doroshenko}, V., {Lutovinov}, A.~A., {Mushtukov}, A.~A., \& {Poutanen}, J. 2017, \aap, 605, A39, \dodoi{10.1051/0004-6361/201730553}

\bibitem[{{Tsygankov} {et~al.}(2006){Tsygankov}, {Lutovinov}, {Churazov}, \& {Sunyaev}}]{Tsygankov_2006}
{Tsygankov}, S.~S., {Lutovinov}, A.~A., {Churazov}, E.~M., \& {Sunyaev}, R.~A. 2006, \mnras, 371, 19, \dodoi{10.1111/j.1365-2966.2006.10610.x}

\bibitem[{{Varun,} {et~al.}(2019){Varun,}, {Maitra}, {Raichur}, \& {Paul}}]{Varun_2019a}
{Varun,}, {Pradhan}, P., {Maitra}, C., {Raichur}, H., \& {Paul}, B. 2019, \apj, 880, 61, \dodoi{10.3847/1538-4357/ab2763}

\bibitem[{{Wang} {et~al.}(2022){Wang}, {Kong}, {Zhang}, {Doroshenko}, {Santangelo}, {Ji}, {Yorgancioglu}, {Chen}, {Zhang}, {Qu}, {Ge}, {Li}, {Chang}, {Tao}, {Peng}, \& {Shui}}]{Wang_2022}
{Wang}, P.~J., {Kong}, L.~D., {Zhang}, S., {et~al.} 2022, \apj, 935, 125, \dodoi{10.3847/1538-4357/ac8230}

\bibitem[{{Xiao} \& {Ji}(2024)}]{Xiao_2024}
{Xiao}, H., \& {Ji}, L. 2024, \apj, 963, 42, \dodoi{10.3847/1538-4357/ad23cd}

\end{thebibliography}
\bibliographystyle{aasjournal}



\end{document}